\documentclass[12pt]{article}

\catcode`\@=11
\@addtoreset{equation}{section}

\global\arraycolsep=2pt
\usepackage{amsbsy,amssymb,latexsym,amsfonts,amsmath}
\usepackage{graphicx,color}

\allowdisplaybreaks

\begin{document}
\begin{flushright}
\parbox{4.2cm}
{}
\end{flushright}

\vspace*{0.7cm}

\begin{center}
{\Large \bf 
Refined topological amplitudes in

  $\mathcal{N}=1$ flux compactification}
\vspace*{2.0cm}\\
{Yu Nakayama}
\end{center}
\vspace*{-0.2cm}
\begin{center}
{\it California Institute of Technology, Pasadena, CA 91125, USA}\footnote{On leave from Berkeley Center for Theoretical Physics at University of California, Berkeley.}
\vspace{3.8cm}
\end{center}

\begin{abstract} 
We study the implication of refined topological string amplitudes in the supersymmetric $\mathcal{N}=1$ flux compactification. They generate higher derivative couplings among the vector multiplets and graviphoton with generically non-holomorphic moduli dependence. For a particular term, we can compute them by assuming the geometric engineering. We claim that the Dijkgraaf-Vafa large $N$ matrix model with the $\beta$-ensemble measure directly computes the higher derivative corrections to the supersymmetric effective action of the supersymmetric $\mathcal{N}=1$ gauge theory.
\end{abstract}

\thispagestyle{empty} 

\setcounter{page}{0}

\newpage

\section{Introduction} 
The flux compactification is ubiquitous in modern applications of string theory.  It is not only generating promising candidates for the realistic particle phenomenology, but also serving as an indispensable tool to understand strong dynamics of the gauge theory through AdS/CFT correspondence and geometric engineering. One of the salient feature of the string compactification is that we can compute the higher derivative corrections that are not accessible to the supergravity construction. Moreover, they are not unrealistic quantities computable only in principle, but they are, under a certain circumstance, realistically computable {\it in practice}.

One typical example is topological string amplitudes. In the physical string theory, they give rise to particular higher derivative corrections to the holomorphic F-terms in the $\mathcal{N}=2$ supersymmetric action \cite{Antoniadis:1993ze}\cite{Bershadsky:1993cx}. In physical applications, they have given important clues in understanding the higher derivative gravitational corrections to the black hole entropy \cite{Ooguri:2004zv}. After the partial supersymmetry breaking via the flux, they also directly compute certain higher derivative corrections in the effective $\mathcal{N}=1$ supersymmetric gauge theories through a geometric engineering.

Topological string theory is a toy model for the physical string theory, yet they compute certain non-renormalized protected quantities in the physical string theory. Quite recently, Antoniadis et al \cite{Antoniadis:2010iq} proposed a vast generalization of the computation in the topological string theory, which we call ``refined topological string amplitudes". They have argued that the refined topological string amplitudes compute new higher derivative corrections to the $\mathcal{N}=2$ supersymmetric effective action in the Calabi-Yau compactification. They have further conjectured that by choosing a suitable class of refined topological amplitudes, the Nekrasov partition function of the $\mathcal{N}=2$ supersymmetric gauge theory with the two parameter extension \cite{Nekrasov:2002qd}\cite{Moore:1997dj}\cite{Losev:1997tp} can be reproduced.\footnote{The Nekrasov partition function with the two parameters is supposed to be related to the refined Gopakumar-Vafa invariant \cite{Gopakumar:1998ii}\cite{Gopakumar:1998jq}\cite{Iqbal:2007ii} and the motivic Donaldson-Thomas invariant \cite{MDT}. We can regard the work of \cite{Antoniadis:2010iq} as the first formulation in terms of the topological string theory.}

The aim of this paper is to discuss the implication of the refined topological string amplitudes in the $\mathcal{N}=1$ supersymmetric flux compactification. It is known that the unrefined topological string amplitudes compute certain F-terms (and their higher derivative corrections) in the flux compactification. We claim that the generalization for the refined topological string amplitudes exist and they compute higher derivative couplings among the vector multiplets and graviphoton with generically non-holomorphic moduli dependence. 

We further claim that these new higher derivative corrections to the effective actions for $\mathcal{N}=1$ supersymmetric gauge theories are computable by using the geometric engineering and the above mentioned conjecture. Indeed a certain class of refined topological amplitudes can be computed by using Nekrasov's localization technique once we assume the conjecture. Furthermore, we propose that the Dijkgraaf-Vafa matrix model \cite{Dijkgraaf:2002fc}\cite{Dijkgraaf:2002dh} with $\beta$-deformed ensemble \cite{Dijkgraaf:2009pc}  directly compute the $\mathcal{N}=1$ supersymmetric higher derivative action.

The organization of the paper is as follows. In section 2, we first review the refined topological string amplitudes in $\mathcal{N}=2$ Calabi-Yau compactification and study the implications in the $\mathcal{N}=1$ supersymmettic flux compactification. In section 3, we verify the $\mathcal{N}=1$ supersymmetric effective action proposed in section 2 by using the $\hat{c}=5$ string formulation. In section 4, we discuss various applications of the refined topological string amplitudes in $\mathcal{N}=1$ supersymmetric compactification and practical methods for their computation. We conjecture  that the Dijkgraaf-Vafa matrix model with $\beta$-deformed ensamble directly compute the $\mathcal{N}=1$ supersymmetric higher derivative action. We also study further breaking of the supersymmetry and generation of the soft terms. In section 5, we conclude the paper with discussions.

\section{Refined topological amplitude reduced to $\mathcal{N}=1$}
Antoniadis et al introduced the ``refined topological amplitudes" that compute certain generalized F-terms in Calabi-Yau compactification of the type II string theory with the $\mathcal{N}=2$ supersymmetry \cite{Antoniadis:2010iq} (see also \cite{Morales:1996bp} for the hetrotic string computation). The generalized higher derivative F-term takes the form of
\begin{align}
S_{g,n} = \int d^4x d^4\theta \mathcal{W}^{2g} \Upsilon^n \ , \label{refined}
\end{align}
where $\mathcal{W}$ is the Weyl superfield that has a component expansion
\begin{align}
\mathcal{W}_{\mu\nu}^{ij} = \epsilon^{ij} T_{\mu\nu}^- - R_{\mu\nu\rho\sigma}^- (\theta^i \sigma^{\rho\sigma} \theta^j) + \cdots \ .
\end{align}
Here $T_{\mu\nu}^-$ is the anti-selfdual graviphoton field strength and $R_{\mu\nu\rho\sigma}^-$ is the anti-selfdual Riemann tensor. $i=1,2$ denotes the $SU(2)_R$-symmetry indices of the $\mathcal{N}=2$ supersymmetry. The superfield $\Upsilon$ denotes the chiral projection of the function of vector multiplets. More precisely, a vector multiplet is described by the reduced $\mathcal{N}=2$ chiral superfield:
\begin{align}
X^I = x^I + \frac{1}{2}F_{\mu\nu}^{+ I}\epsilon_{ij}(\theta^i \sigma^{\mu\nu} \theta^j) + \cdots \ ,
\end{align}
where $I=0, \cdots h$, and $F^{+ I}_{\mu\nu}$ are selfdual field strength for the $I$th vector field ($I=0$ will be identified with the graviphoton while $I=1,\cdots h$ will be in the vector multiples).
 The chiral projected superfield is given by
\begin{align}
\Upsilon = \Pi\left( \frac{G(X^I/X^0, (X^I/X^0)^\dagger)}{(X^0)^2} \right)\ ,
\end{align}
where $\Pi = (\epsilon_{ij} \bar{D}^i \bar{\sigma}_{\mu\nu} \bar{D}^j)^2$ is the superconformal chiral projection operator (see e.g. \cite{deRoo:1980mm}). For later purposes we will introduce the physical moduli field as  $\phi^I = X^I/X^0$ ($I=1,\cdots, h$). $X^0$ plays the role of the compensator, and in the rigid limit, we often fix the superconformal gauge so that $X^0 = \text{const}$ in the following.
 Note that we can define $\Upsilon$ for {\it any} possibly non-holomorphic function $G$, so the amplitudes \eqref{refined} have huge degrees of freedom compared with the $n=0$ holomorphic amplitude. With respect to this arbitrariness, one should regard $\Upsilon^n$ as an abbreviation of $\prod_{i=1}^n\Upsilon_{G_i}$ because one can choose different $\Upsilon$ at each order.

They showed \cite{Antoniadis:2010iq} that the amplituds \eqref{refined}
\begin{align}
S_{g,n} = \int d^4x F_{g,n}(\phi,\bar{\phi})(R^-)^2 (T^-)^{2g-2}(F^+)^{2n} + \cdots  
\end{align}
can be computed within the topological string theory. A direct computation in the type II string theory \cite{Antoniadis:2010iq} relates the higher derivative F-terms with the topological string amplitudes:
\begin{align}
F_{g,n} = \int_{\mathcal{M}_{(g,n)}} \langle \prod_{k=1}^{3g-3+n} |(\mu_k \cdot G^{-})|^2 \prod_{k=1}^{n} \int \Psi_{I_k} \prod_{l=1}^n \hat{\Psi}_{J_l}\rangle_{\text{top}} \ . \label{refinedt}
\end{align}
Here, $\Psi_I$ are (anti-chiral, chiral) primary operators with the $U(1)$ $R$-charge $(-1,1)$ and twisted conformal dimension $(1,1)$, which will be integrated over the Riemann surface. The hatted operators are defined as $\hat{\Psi}_{J} = \oint dz \rho(z) \oint d\bar{z} \tilde{\rho}(\bar{z}) \Psi_J$, where $\rho$ is the unique left-moving operator with the charge $+3$ and dimension $0$. Thus, $\hat{\Psi}_J$ have $U(1)$ $R$-charge $(+2,-2)$ and dimension $(0,0)$, and they are in the twisted BRST cohomology. They are located at $n$ distinct punctures on the genus $g$ Riemann surface.  $\mu_k$ is the Beltrami-differential associated with the complex structure moduli space $\mathcal{M}_{(g,n)}$ for the $n$-punctured genus $g$ Riemann surfaces. The natural measure for the A-twist is given by
\begin{align}
|(\mu_k \cdot G^{-})|^2 = (\mu_k \cdot{G}^-)(\bar{\mu}_k\cdot \bar{G}^+)  \ .
\end{align}

When $n=0$, the amplitude corresponds to the partition function of the genus $g$ topological string theory, and it computes the graviphoton corrections to the $\mathcal{N}=2$ prepotential. It is important to observe that the amplitude $F_{g,n}$ for $n \ge 1$ is not holomorphic because the insertion $\int \Psi_{I_k}$ is not necessarily annihilated by the topological BRST charge.

Furthermore, \eqref{refinedt} should be regularized so that the leading $O(1/|z|^2)$ singularities associated with the OPE $\Psi(z)_I \hat{\Psi}(0)_J$ are subtracted. This is physically necessary because one has to compute the 1PI diagram rather than the S-matrix to obtain the effective action. The regularization procedure with the subtraction will remove the non-1PI diagrams appearing in  the external lines that will generate ``massless" one particle state interior. It is interesting to observe that from the charge assignment, once one subtracts the leading order singularity, the first non-trivial OPE between $\Psi_I$ and $\hat{\Psi}_J$ should give birth to the operator whose $R$-charge is $(1,1)$ and the twisted conformal dimension $(0,0)$. Unless it is given by a certain descendent operator, it has the same charge as the topological chiral primary operator, and the amplitude is likely to be holomorphic (up to a possible holomorphic anomaly) \cite{Nakayama:2010ff}. We will come back to this point later in section 3.

The main goal of this paper is to study the $\mathcal{N}=1$ reduction of the refined topological amplitudes. In the context of the type II string theory, we will introduce the background fluxes to break the $\mathcal{N}=2$ supersymmetry down to $\mathcal{N}=1$. Within the holomorphic topological string theory, the effect of the flux and subsequent supersymmetry breaking was first studied in \cite{Vafa:2000wi} by using the so-called spurion method.

For definiteness, let us focus on the type IIA compactification. The idea is to substitute the background superfield
\begin{align}
t^I = t^I + N^I \Theta^2 \label{repl}
\end{align}
into the effective action to break the $\mathcal{N}=2$ supersymmetry by the explicit vacuum expectation value corresponding to the $\mathcal{N}=1$ supersymmetry preserving flux and read off the $\mathcal{N}=1$ effective action after the partial supersymmetry breaking.\footnote{Technically, the special coordinates $\phi^I$ and the large radius geometric K\"ahler moduli $t^I$ are related by a non-trivial coordinate transformation. We will be implicit about the detail of the transformation in the following.}
Here $\Theta$ denotes a particular component of the supersymmetry which will be broken by the flux, and the unbroken $\mathcal{N}=1$ supercoordinate will be denoted by $\vartheta$. $N^I$ is the number of R-R flux through the two-cycle corresponding to the K\"ahler moduli $t^I$, but it also acquires the imaginary part coming from the NS-NS flux \cite{Vafa:2000wi}. 
As discussed in \cite{Vafa:2000wi}, for $n=0$, the conventional holomorphic (unrefined) topological string amplitudes give the (graviphoton corrected) $\mathcal{N}=1$ superpotential term:
\begin{align}
W_g = \int d^2\vartheta (\mathcal{W}^-)^{2g} \left( N^I \frac{\partial F_g}{\partial t^I}\right) \ , \label{grvf}
\end{align}
where $\mathcal{W}^-$ is the $\mathcal{N}=1$ reduced graviphoton superfield.

We claim that we can compute the effective $\mathcal{N}=1$ action reduced from the $\mathcal{N}=2$ refined topological amplitude in the similar manner. For this purpose, as in the $\mathcal{N}=2$ case, we introduce chiral projection of the general $\mathcal{N}=1$ superfield: $U = \bar{\mathcal{D}}^2 F(X,X^\dagger, W , \bar{W})$ so that $\bar{\mathcal{D}} U = 0$.
Here, we have decomposed an $\mathcal{N}=2$ vector multiplet into an $\mathcal{N}=1$ chiral multiplet $X$ and an $\mathcal{N}=1$ vector multiplet $V$ whose field strength is $W_{\alpha} = \bar{\mathcal{D}}^2 \mathcal{D}_{\alpha} V$, and the $\mathcal{N}=1$ superderivative is denoted by $\mathcal{D}$.
With this notation, the $\mathcal{N}=1$ amplitudes that can be reduced from the $\mathcal{N}=2$ amplitudes fall in the class of 
\begin{align}
W_{g,n} = \int d^2 \vartheta (\mathcal{W^-})^{2g} U^n \ . \label{ref}
\end{align}
As in the $\mathcal{N}=2$ case, we should regard $U^n$ as an abbreviation of $\Pi_{i=1}^n U_{F_i}$ because we can choose different $U$ at each order.

In contrast to its appearance, the amplitude \eqref{ref} is not holomorphic because the ``chiral" superfield $U$ contains arbitrary non-holomorphic dependence through $F(X,X^\dagger,W,\bar{W})$. Indeed, the ``F-term" amplitude \eqref{ref} can be rewritten as a ``D-term":
\begin{align}
W_{g,n} = \int d^2 \vartheta d^2\bar{\vartheta} (\mathcal{W^-})^{2g} F(X,{X}^\dagger, W, \bar{W}) U^{n-1} \label{red0}
\end{align}
We note that the existence of $F$ may be only local, and there might be a global obstruction. In that case, the ``F-term" in \eqref{ref} cannot be integrated to the D-term \eqref{red0}. This is the phenomenon closely related to the one found in the context of the ``higher derivative F-term" in \cite{Beasley:2004ys}.

As in the genuine superpotential term \eqref{grvf}, we claim that the replacement \eqref{repl} in the $\mathcal{N}=2$ effective action is a good approximation to compute the $\mathcal{N}=1$ generalized ``higher derivative F-term" as long as the supersymmetry breaking is small and the back-reaction to the geometry is small. In the extreme warped case, the D-terms may acquire significant corrections and the extra contributions other than the refined topological string amplitudes cannot be neglected.

Accordingly, with the substitution \eqref{repl}, the refined topological string amplitude \eqref{refined} gives birth to the $\mathcal{N}=1$ effective action \eqref{ref} with one $U$ given by $ N^I \bar{\mathcal{D}}^2 \left(\frac{\partial^3 G(\phi,\phi^\dagger)}{\partial t^I\partial \phi^{J\dagger}  \partial \phi^{K\dagger}}\bar{W}^J\bar{W}^K \right)$ and the other $(n-1)$ $U$'s given by $\bar{\mathcal{D}}^2 \left( \frac{\partial^2 G(\phi,\phi^\dagger)}{\partial \phi^{J\dagger} \partial \phi^{K\dagger}} \bar{W}^J\bar{W}^K\right)$ so that
\begin{align}
W_{g,n} &= \int d^2 \vartheta (\mathcal{W}^-)^{2g} N^I \bar{\mathcal{D}}^2 \left(\frac{\partial^3 G(\phi,\phi^\dagger)}{\partial {t^I} \partial \phi^\dagger \partial \phi^\dagger}\bar{W}^2 \right) \left(\bar{\mathcal{D}}^2 \left( \frac{\partial^2 G(\phi,\phi^\dagger)}{\partial \phi^\dagger \partial \phi^\dagger} \bar{W}^2\right)\right)^{n-1} \cr
& =  \int d^2 \vartheta d^2 \bar{\vartheta} (\mathcal{W}^-)^{2g} N^I \left(\frac{\partial^3 G(\phi,\phi^\dagger)}{\partial t^I \partial \phi^\dagger \partial \phi^\dagger}\bar{W}^2 \right) \left(\bar{\mathcal{D}}^2 \left( \frac{\partial^2 G(\phi,\phi^\dagger)}{\partial \phi^\dagger \partial \phi^\dagger} \bar{W}^2\right)\right)^{n-1} \ . \label{n1}
\end{align}
We emphasize that the precise detail of $U$, or $G$ on the moduli dependence can be computed from the refined topological amplitudes \eqref{refinedt}.
 In the next section, we discuss the validity of the simple replacement \eqref{repl} to break the supersymmetry in the $\hat{c}=5$ string formulation.

\section{$\hat{c}=5$ string formulation for refined amplitude}
To understand the effect of the partial supersymmetry breaking from $\mathcal{N}=2$ to $\mathcal{N}=1$ in the flux compactification, it is convenient to use the $\hat{c}=5$ string formulation as used in establishing the relation between the unrefined topological string amplitudes and the $\mathcal{N}=1$ supersymmetric effective action in \cite{Berkovits:2003pq}. 

The $\mathcal{N}=2$, $\hat{c} = 5$ string theory has the worldsheet action
\begin{align}
S = \int d^2z (p_1^\alpha \bar{\partial}\theta^1_\alpha  + p_2^\alpha \partial \theta^2_\alpha + \frac{1}{2} \epsilon_{\alpha\beta} \partial x^{\alpha \dot{+}} \bar{\partial} x^{\beta \dot{-}}) + S_{\mathrm{CY}} \ ,
\end{align}
where the Calabi-Yau part of the action is topologically twisted. $\theta^{i}$ and $p^i$ are Grassmann-valued fields.
The theory has the following twisted $\mathcal{N}=2$ generators:
\begin{align}
T &= p_{1\alpha} \partial \theta^{1\alpha} 
+ \frac{1}{2} \epsilon_{\alpha \beta} \partial x^{\alpha \dot{+}} \partial x^{\beta \dot{-}} + T_{\mathrm{CY}} \cr 
G^+ &= \theta_{1\alpha} \partial x^{\alpha \dot{+}}  + G^+_{\mathrm{CY}} \ , \ \ G^- = p_{1\alpha} \partial x^{\alpha \dot{-}}  + G^-_{\mathrm{CY}} \cr
J &= \theta_{1}^{\alpha} p^{1}_{\alpha} + J_{\mathrm{CY}} \ \cr
\bar{T} &= p_{2\alpha} \partial \theta^{2\alpha} 
+ \frac{1}{2} \epsilon_{\alpha \beta} \bar{\partial} x^{\alpha \dot{+}} \bar{\partial} x^{\beta \dot{-}} + \bar{T}_{\mathrm{CY}} \cr 
\bar{G}^- &= \theta_{2\alpha} \bar{\partial} x^{\alpha \dot{+}}  + \bar{G}^-_{\mathrm{CY}} \ , \ \ \bar{G}^+ = p_{2\alpha} \bar{\partial} x^{\alpha \dot{-}}  + \bar{G}^+_{\mathrm{CY}} \cr
\bar{J} &= \theta_{2}^{\alpha} p^{2}_{\alpha} + \bar{J}_{\mathrm{CY}} \ .
\end{align}
Here, $\{T_{\mathrm{CY}},G^{+}_{\mathrm{CY}}, G^{-}_{\mathrm{CY}}, J_{\mathrm{CY}}\}$ are the twisted $\mathcal{N}=2$ generators of the internal Calabi-Yau conformal field theory. See \cite{Berkovits:2003pq} for the detailed structure of the $\hat{c}=5$ string theory.\footnote{We use a slightly different notation for the superspace so that $(\theta, \bar{\theta})$ there is replaced by $(\theta_1,\theta_2)$ here. Also our $\bar{G}^+$ is their $\bar{G}^-$ and vice versa.}

In this paper, we would like to compute the ``refined" $\hat{c} = 5$ string amplitude:
\begin{align}
A_{g,M,n} = \left\langle |\prod_{j=1}^{3g-3 + M+n} \int dm^j\int \mu_j G^-|^2 \prod_{r=1}^{M} V_{r}(z_r) \prod_{k=1}^{n} \int \tilde{\Upsilon}_{I_k}\Psi_{I_k} \prod_{l=1}^n \hat{\Upsilon}_{J_l} \hat{\Psi}_{J_l} \right\rangle  \ .
\end{align}
We focus on the case where $M=2g$ and all $V_{r}$ are given by the graviphoton vertex operators. In addition, we choose $2g$ of the Beltrami differentials to be associated with the positions of the graviphoton vertex operators. This leads to 
\begin{align}
A_{g,n} = \left\langle |\prod_{j=1}^{3g-3+n} \int dm^j\int \mu_j G^-|^2 \prod_{r=1}^{2g} \int d^2z_r W_{r}(z_r) \prod_{k=1}^{n} \int \tilde{\Upsilon}_{I_k}\Psi_{I_k} \prod_{l=1}^n \hat{\Upsilon}_{J_l} \hat{\Psi}_{J_l} \right\rangle  \ .
\end{align}
where $W_r =  \oint G^{-} \oint \bar{G}^+ R$ is the graviphoton superfield.

The refined $\hat{c}=5$ amplitudes are somewhat more subtle than those considered in \cite{Berkovits:2003pq}. First of all, as we mentioned in section 2, in the computation of the twisted correlation function, one has to subtract the leading order $1/|z|^2$ singularities associated with the OPE $\Psi_I(z) \hat{\Psi}_J(0)$. Correspondingly, note that the insertion by $\Psi_I$ is not necessarily BRST invariant, so the amplitude is not generically holomorphic. Our motivation to study the generically non-BRST invariant $\hat{c}=5$ amplitude is to reproduce the $\mathcal{N}=2$ RNS string theory refined amplitudes discussed in section 2. For this purpose, the superfield vertex operators $\tilde{\Upsilon}$ and $\hat{\Upsilon}$ satisfy a particular relation as we will see.

We can integrate over the zero modes of $(x,\theta^i)$ and we obtain the refined topological amplitudes in $\hat{c}=5$ string theory formulation:
\begin{align}
A_{g,n} =& \int d^4 x d^4\theta \mathcal{W}^{2g} \Upsilon^n \cr
           &\times \left\langle |\prod_{j=1}^{3g-3+n} \int dm^j\int \mu_j G^-|^2  \prod_{k=1}^{n} \int \Psi_{I_k} \prod_{l=1}^n \hat{\Psi}_{J_l} \right\rangle_{\mathrm{top}} \ . 
\end{align}
Here, for the consistency of the amplitude and supersymmetry, the superfield $\Upsilon = \tilde{\Upsilon}\hat{\Upsilon}$ must be a chiral superfield. We further demand that it is given by the chiral projection of the function of the vector multiplets\footnote{Strictly speaking, this information cannot be encoded in the $\hat{c}=5$ formulation because there is no notion of the $ \bar{\theta}^{\dot{\alpha}}$ variables. We regard it as an extra input.}
\begin{align}
\Upsilon = \Pi\left( \frac{G(X^I/X^0, (X^I/X^0)^\dagger)}{(X^0)^2} \right)\ .
\end{align}
The coefficient in the effective action is precisely the refined topological string amplitudes discussed in section 2. In this way, we have reproduced the $\mathcal{N}=2$ refined type II RNS string amplitudes in the $\hat{c}=5$ formalism.

We now break the supersymmetry to compute the $\mathcal{N}=1$ effective action  by introducing the expectation value for the auxiliary fields in $\Upsilon = \partial^2 G(\phi,\phi^\dagger)/(\partial \phi^\dagger \partial \phi^\dagger) \bar{W}^2 + D^2 \Theta^2 + \cdots$. More precisely, we demand that the supersymmetry breaking is induced by the expectation value for the auxiliary chiral superfield corresponding to the harmonic two-form (i.e. geometric K\"ahler moduli): $t^I = t^I +  N^I \Theta^2 $, where $N^I$ is the corresponding $\mathcal{N}=1$ preserving flux. Note that the specification of the supersymmetry breaking is not directly dictated in the $\hat{c}=5$ formulation, so it is in principle possible to study broader classes of the supersymmetry breaking than the $\mathcal{N}=1$ partial supersymmetry breaking we are focusing here (see section 4.3 for further discussions). 
As a result of the partial supersymmetry breaking from the flux, we obtain the $\mathcal{N}=1$ supersymmetric effective action
\begin{align}
S_{g,n} &= \int d^4x d^2 \vartheta (\mathcal{W^-})^{2g} N^I \bar{\mathcal{D}}^2 \left(\frac{\partial^3 G(\phi,\phi^\dagger)}{ \partial t^I \partial \phi^\dagger \partial \phi^\dagger}\bar{W}^2 \right) \left(\bar{\mathcal{D}}^2 \left( \frac{\partial^2 G(\phi,\phi^\dagger)}{\partial \phi^\dagger \partial \phi^\dagger} \bar{W}^2\right)\right)^{n-1} \ ,  \label{n2}
\end{align}
which is the same effective action we proposed in the previous section.

\section{Applications}
In this section, we discuss several applications and explicit examples of the refined topological amplitudes and their applications in the $\mathcal{N}=1$ supersymmetric flux compactification.

\subsection{Universal contributions from conifold singularity}
As an example of the refined topological amplitudes and their contribution to the $\mathcal{N}=1$ supersymmetric effective action in a flux compactification, let us consider the conifold singularity. The conifold singularity is rather ubiquitus in string compctifications and it gives a universal contribution to the $\mathcal{N}=2$ (refined) amplitudes as well as the corresponding $\mathcal{N}=1$ supersymmetric amplitudes in the flux compactification. 

The refined topological string amplitudes for the conifold was computed in \cite{Nakayama:2010ff} in agreement with the matrix model proposal in \cite{Dijkgraaf:2009pc}. 
For the $\mathcal{N}=2$ supersymmetric amplitude, it is custumary to introduce the generating function
\begin{align}
F(\epsilon_-, \epsilon_+) = \sum_{g=1}^\infty \sum_{n=1}^\infty \epsilon_-^{2g} \epsilon_+^{2n} F_{g,n} \ .
\end{align}
The direct computation gives
\begin{align}
F \sim \int \frac{dt}{t} \frac{\pi \epsilon_1}{\sin(\pi \epsilon_1 t)} \frac{\pi \epsilon_2}{\sin(\pi \epsilon_2 t)} e^{-t \mu} \ , \label{opat}
\end{align}
where $\epsilon_1 = \epsilon_+ + \epsilon_-$ and $\epsilon_2 = \epsilon_+ -\epsilon_-$. The amplitude is nothing but the free energy of the $c=1$ non-critical string theory at the radius $\sqrt{2}|\epsilon_1/\epsilon_2|$ \cite{Gross:1990ub} (consult e.g. \cite{Nakayama:2004vk} for the reason why the non-critical string appears in the conifold). 

The corresponding $\mathcal{N}=1$ supersymmetric amplitude is given by
\begin{align}
S_{g,n} &= \int d^4x \int d^2 \vartheta N\left( \frac{\partial F_{g,n}}{\partial \mu} (\mathcal{W}^-)^{2g} (\bar{\mathcal{D}}^2 \bar{W}^2)^n \right) \ \cr
&= \int d^4x \int d^2 \vartheta d^2\bar{\vartheta} N\left( \frac{\partial F_{g,n}}{\partial \mu} (\mathcal{W}^-)^{2g} \bar{W}^2 (\bar{\mathcal{D}}^2 \bar{W}^2)^{n-1} \right) 
\end{align}
Note that in the case of conifold, $F_{g,n}$ is actually {\it holomorphic} in $\mu$, so one can commute $\bar{\mathcal{D}}^2$ in $U$ with the holomorphic function (or more precisely a section of the suitable line bundle) $F_{g,n}(\mu)$. 

For instance, the lowest correction comes from the ``genus $1$" part of the refined topological amplitudes: $F_1 \sim (\epsilon_1^2 + \epsilon_2^2) \log \mu = \frac{1}{2}(\epsilon_+^2 + \epsilon_-^2) \log \mu$. The anti-selfdual part gives the graviphoton corrected superpotential term
\begin{align}
S_{1,0} = \int d^4 x \int d^2\vartheta \frac{N}{2} (\mathcal{W}^-)^2 \mu^{-1} \ 
\end{align}
while the selfdual part gives the generalized F-term introduced in section 2:
\begin{align}
S_{0,1} &= \int d^4x\int d^2 \vartheta \frac{N}{2}(\bar{\mathcal{D}}^2 \bar{W}^2) \mu^{-1} \cr
        &=  \int d^4x\int d^2 \vartheta d^2 \bar{\vartheta} \frac{N}{2}(\bar{W}^2) \mu^{-1} \ .
\end{align}
On the Minkowski vacuum, the vacuum expectation value of $\mu$ is determined from the tree level part of the flux superpotential, which gives us the famous gaugino condensation effect $\mu \sim N \Lambda^3$ irrespective of the higer derivative corrections newly discussed in this paper.

It is interesting to observe that the $\mathcal{N}=2$ topological string amplitude considered here possesses a symmetry between $\epsilon_+$ and $\epsilon_-$ (as well as between $\epsilon_1$ and $\epsilon_2$). The corresponding $\mathcal{N}=1$ supersymmetric action has the resultant symmetry under the exchange of the graviphoton field strength $T^{-}$ and the vector multiplet field strength $F^{+}$. The first symmetry is only true when the BPS spectrum of the $\mathcal{N}=2$ theory has a symmetry under the left spin and the right spin.
 The latter symmetry is due to the T-duality of the compactified boson in the $c=1$ non-critical string theory language.\footnote{The symmetry between $\epsilon_1$ and $\epsilon_2$ naturally arises in the Nekrasov partition function due to  a trivial symmetry of the exchange of $x_{12}$ plane and $x_{34}$ plane in $\mathbf{R}^4$. The symmetry between $\epsilon_+$ and $\epsilon_-$ is not manifest because the R-symmetry twist does not preserve it.}

\subsection{$\beta$-deformed matrix model conjecture}
While it is not always feasible to compute the refined topological string amplitudes based on the first principle from their definition except for the simplest conifold case, there exist several tricks to compute them indirectly by using  duality arguments. Our basic idea is based on the conjecture \cite{Antoniadis:2010iq} that the refined topological string amplitude for a particular choice of the vector multiplet is equivalent to the $\Omega$ deformation in the Nekrasov partition function of the $\mathcal{N}=2$ supersymmetric gauge theory.

The cojecture is applicable when the target space Calabi-Yau has a geometric-engineering interpretation from a gauge theory. Then it is well-established that the (unrefined) topological string partition function yields the Nekrasov partition function by identifying the graviphoton correction with the Nekrasov twist parameter with $\epsilon_1 = -\epsilon_2$. It is conjectured that the refined topological string partition function computes the Nekrasov partition function with the more general parameter $\epsilon_1 \neq -\epsilon_2$. Since the Nekrasov partition function itself can be computed by the localization technique for general values of $\epsilon_1$ and $\epsilon_2$, we can compute $F_{g,n}$ and the corresponding $\mathcal{N}=1$ supersymmetric effctive action that we have discussed from the conjecture. In the simplest conifold case, the conjecture is verified by a direct computation in \cite{Nakayama:2010ff}.

An alternative and slightly more direct way to compute the refined topological amplitues, again for a particular choice of the vector multiplet, is to use the large $N$ matrix model partition function with the so-called $\beta$-ensemble. For a particular type of Calabi-Yau singularity, the unrefined topological string partition function can be computed  by the large $N$ (Dijkgraaf-Vafa) matrix integral \cite{Dijkgraaf:2002fc}\cite{Dijkgraaf:2002dh}:
\begin{align}
Z = \int D\Phi e^{-V(\Phi)} \ , \label{matrix}
\end{align}
where $\Phi$ is the $N\times N$ Hermitian matrix and $V(\Phi)$ encodes the geometry of the Calabi-Yau. The matrix integral measure is the conventional Haar measure (divided by the volume of the $U(N)$ group).
We identify the $1/N$ expansion as the topological string expansion. In the simplest conifold case the potential is $V(\Phi) = \frac{\mu}{2} \mathrm{Tr}\Phi^2$. 

It was argued in \cite{Dijkgraaf:2009pc} that the Nekrasov deformation can be introduced by changing the measure of the matrix integral \eqref{matrix}. When we diagonalize the Hermitian matrix $\Phi$, the original measure is given by the Vandermode determinant of the eigenvalues:
\begin{align}
Z = \int d^N z \prod_{I<J} (z_I-z_J)^2 e^{-V(z)} \ .
\end{align}
To compute the refined topological amplitude, we replace the Vandermode measure by the so-called $\beta$-ensemble measure:
\begin{align}
Z_{\beta} = \int d^N z \prod_{I<J} (z_I-z_J)^{2\beta} e^{-V(z)} \ .
\end{align}
We identify $\beta = -\frac{\epsilon_1}{\epsilon_2}$. 

As demonstrated in \cite{Dijkgraaf:2009pc}, for the simple Gaussian potential corresponding to the conifold geometry, the $\beta$-ensemble integral gives the desired refined topological string amplitudes in the large $N$ limit:
\begin{align}
\log Z_{\beta} = \int \frac{ds}{s}\frac{e^{\mu s}}{(1-e^{\epsilon_1s})(1-e^{\epsilon_2s})} \ .
\end{align}
The expansion with respect to $1/\mu$ gives the universal contribution discussed in section 4.1. More generally, the matrix model with the $\beta$-ensemble for the Nekrasov partition function was studied in \cite{Sulkowski:2009ne}.

In this way, one can interpret the partition function of the matrix integral with the $\beta$-ensemble measure directly in terms of the $\mathcal{N}=1$ theory. In tern, the geometric engineering gives us the way to compute the higher derivative F-terms in low-energy $\mathcal{N}=1$ supersymmetric gauge theory from the Dijkgraaf-Vafa large $N$ matrix model with the $\beta$-ensemble measure. In the original unrefined Dijkgraaf-Vafa conjecture, the $\mathcal{N}=1$ supersymmetric Yang-Milles theory (with adjoint matter) realized by D5-brane wrapping singular two-cycles of the Calabi-Yau is replaced by the flux geometry, whose effective action is computed by the topological string theory and identified with the low-effective action of the $\mathcal{N}=1$ supersymmetric gauge theory with the gaugino condensation.
The refined generalization of us leads to the following conjecture:

{\bf Conjecture}

{\it The higher derivative F-terms \eqref{n1} appearing in the low energy effective action of the $\mathcal{N}=1$ supersymmetric $U(n)$ gauge theory with adjoint matter can be computed by the Dijkgraaf-Vafa large $N$ matrix model with the $\beta$-deformed ensemble.}

In this conjecture, we interpret the chiral multiplet appearing in the higher derivative F-terms by the gaugino condensate $S \sim \mathrm{Tr}W^2$. The selfdual field strength appearing in the $\mathcal{N}=1$ supersymmetric effective action may be interpreted as the $U(1)$ vector multiplet in the $U(n)$ gauge theory.

The refined topological amplitude computed from the $\beta$-deformed matrix integral is holomorphic with respect to the moduli fields. In this sense, the refined amplitude computed in this way is very specific. It is also clear from the Nekrasov partition function viewpoint because the Nekrasov partition function has only one extra parameter compared with the unrefined topological amplitude. The refined topological amplitudes studied in section 2 is more generic: it is not necessarily holomorphic and it has as many parameters as the number of vector multiplets. It is of interest to study how one can compute the more general refined amplitudes beyond their original definition.

\subsection{Further supersymmetry breaking}
In section 3, we have studied the partial breaking of the $\mathcal{N}=2$ supersymmetry to $\mathcal{N}=1$ supersymmetry by giving the expectation value to the $\Theta^2$ component of the $\mathcal{N}=2$ superfield as flux contribution. Nothing there forbids us from breaking supersymmetry further down to $\mathcal{N}=0$ supersymmetry by giving an expectation value to the remaining $\mathcal{N}=1$ superfield. The resulting effective action acquires a so-called soft terms via the supersymmetry breaking.

As a typical example, let us study the $\mathcal{N}=1$ $g=0,n=1$ amplitude:
\begin{align}
S_{0,1} = \int d^4 xd^2\vartheta d^2\bar\vartheta \bar{W}^2 K_{0,1}(\Phi,{\Phi}^\dagger, W, \bar{W}) \ .
\end{align}
By setting $\Phi = \phi + \vartheta^2 \langle F\rangle $, we obtain the gaugino mass term:
\begin{align} 
S_{0,1} = \int d^4 x (\bar{\lambda} \bar{\lambda}) \frac{\partial^2 K_{0,1}}{\partial\Phi \partial \Phi^\dagger} |\langle F \rangle|^2 + \cdots \ , 
\end{align}
where $\bar{\lambda}_{\dot{\alpha}}$ is the gaugino which is the lowest component of $\bar{W}_{\dot{\alpha}}$. 

Similarly, there would be a mixed D-term and F-term effects from the $g=0,n=2$ refined amplitudes
\begin{align}
S_{0,2} = \int d^4 xd^2\vartheta d^2\bar\vartheta \bar{W}^2 \bar{\mathcal{D}}^2 \bar{W}^2 K_{0,2}(\Phi,{\Phi}^\dagger, W, \bar{W}) \ ,
\end{align}
which gives after setting $\Phi = \phi + \vartheta^2 \langle F\rangle$ and $\bar{W}^2 = \bar{\lambda}^2 + \bar{\vartheta}^2 \langle D \rangle^2$
\begin{align}
S_{0,2} = \int d^4 x(\bar{\lambda}\bar{\lambda}) \langle D \rangle^2 |\langle F\rangle|^2  \frac{\partial^2 K_{0,2}}{\partial\Phi \partial \Phi^\dagger}  \ .
\end{align}
Such a contribution to the gaugino mass term is important in understanding e.g. strongly coupled D-term gauge mediation \cite{Nakayama:2007cf}\cite{Nakayama:2007je} (or more general gauge mediation \cite{Meade:2008wd}) in supersymmetric particle phenomenology.

In the simplest conifold example, the refined amplitude is holomorphic so there is no gaugino mass term from $S_{0,1}$. Instead, it gives a threshold correction to the gauge coupling constant
\begin{align}
S_{0,1} &= \int d^4 x d^2\vartheta d^2\bar{\vartheta} \bar{W}^2 \frac{1}{S} \cr
        & = -\int d^4 x d^2 \bar{\vartheta} \bar{W}^2 \frac{\langle F\rangle}{S^2}  + \cdots \cr
	& = -\int d^4 x \frac{\langle F\rangle}{S^2}\left( (F^+_{\mu\nu})^2 + \bar{\lambda} \sigma^{\mu} \partial_{\mu} \bar{\lambda} +\cdots \right) + \cdots \ .
\end{align}
Similarly, the mixed D-term and F-term supersymmetry breaking gives a further threshold correction from $S_{0,2}$ as
\begin{align}
S_{0,2} &= \int d^4 x d^2 \vartheta d^2 \bar{\vartheta} \bar{W}^2 \bar{\mathcal{D}}^2 \bar{W}^2 \frac{1}{S^3} \cr
& = -6 \int d^4 x \frac{\langle F \rangle \langle D\rangle^2}{S^4} \left( (F^+_{\mu\nu})^2 + \bar{\lambda} \sigma^{\mu} \partial_{\mu} \bar{\lambda} +\cdots \right) + \cdots \ .
\end{align}
We emphasize the effect is potentially large because $S$, which appears in the denominator, takes an exponentially smaller value compared with the Plank (string) scale due to the effect of the warping (or dimensional mutation in the language of the dual gauge theory).

\section{Discussion}
In this paper we have studied the implications of the refined topological amplitudes in the $\mathcal{N}=1$ supersymmetric flux compactification. We have shown that the refined topological amplitudes compute certain higher derivative F-terms in the effective action. The refined amplitude is generically non-holomorphic, but we have argued that they can be computed by using the geometric engineering and the Dijkgraaf-Vafa large $N$ matrix model with $\beta$-ensemble measure for a certain specific refined amplitude.

Which kind of refined amplitudes are computable? We have noted that in both the Nekrasov localization computation through the geometric engineering or the Dijkgraaf-Vafa large $N$ matrix model computation, the refined topological amplitudes computed are all holomorphic. The requirement of the holomorphicity in the refined topological amplitude seems relevant in finding ``computable" $\mathcal{N}=1$ supersymmetric higher derivative F-terms. In \cite{Nakayama:2010ff}, we have discussed that the regularization by subtracting the singularities play an important role to assure the holomorphicity of the refined topological amplitudes. It would be interesting to study the structure further both from the worldsheet viewpoint as well as the target space effective field theory viewpoint.

The formulation we have discussed in this paper by itself applies to the non-holomorphic amplitude. In particular, the generation of the gaugino mass in the $\mathcal{N}=0$ supersymmetry breaking needs the non-holomorphic dependence on the moduli fields. It would be important to establish ways to compute the non-holomorphic dependence. We expect that the refined topological amplitude must satisfy certain integrability conditions as well as holomorphic anomaly equations, and we believe they will give us a breakthrough in this direction.

\section*{Acknowledgements}
The author would like to thank C.~Beem for discussions.
The work was supported in part by the National Science Foundation under Grant No.\ PHY05-55662 and the UC Berkeley Center for Theoretical Physics.


\begin{thebibliography}{99}
\bibitem{Antoniadis:1993ze}
  I.~Antoniadis, E.~Gava, K.~S.~Narain and T.~R.~Taylor,
  Nucl.\ Phys.\  B {\bf 413}, 162 (1994)
  [arXiv:hep-th/9307158].

\bibitem{Bershadsky:1993cx}
  M.~Bershadsky, S.~Cecotti, H.~Ooguri and C.~Vafa,
  Commun.\ Math.\ Phys.\  {\bf 165}, 311 (1994)
  [arXiv:hep-th/9309140].

\bibitem{Ooguri:2004zv}
  H.~Ooguri, A.~Strominger and C.~Vafa,
  Phys.\ Rev.\  D {\bf 70}, 106007 (2004)
  [arXiv:hep-th/0405146].

\bibitem{Antoniadis:2010iq}
  I.~Antoniadis, S.~Hohenegger, K.~S.~Narain and T.~R.~Taylor,
  Nucl.\ Phys.\  B {\bf 838}, 253 (2010)
  [arXiv:1003.2832 [hep-th]].

\bibitem{Nekrasov:2002qd}
  N.~A.~Nekrasov,
  Adv.\ Theor.\ Math.\ Phys.\  {\bf 7}, 831 (2004)
  [arXiv:hep-th/0206161].
\bibitem{Moore:1997dj}
  G.~W.~Moore, N.~Nekrasov and S.~Shatashvili,
  Commun.\ Math.\ Phys.\  {\bf 209}, 97 (2000)
  [arXiv:hep-th/9712241].

\bibitem{Losev:1997tp}
  A.~Losev, N.~Nekrasov and S.~L.~Shatashvili,
  Nucl.\ Phys.\  B {\bf 534}, 549 (1998)
  [arXiv:hep-th/9711108].

\bibitem{Gopakumar:1998ii}
  R.~Gopakumar and C.~Vafa,
  arXiv:hep-th/9809187.
\bibitem{Gopakumar:1998jq}
  R.~Gopakumar and C.~Vafa,
  arXiv:hep-th/9812127.
\bibitem{Iqbal:2007ii}
  A.~Iqbal, C.~Kozcaz and C.~Vafa,
  JHEP {\bf 0910}, 069 (2009)
  [arXiv:hep-th/0701156].
\bibitem{MDT}
M.~Kontsevich, Y.~Soibelman,
arXiv:0811.2435[math].

\bibitem{Dijkgraaf:2002fc}
  R.~Dijkgraaf and C.~Vafa,
  Nucl.\ Phys.\  B {\bf 644}, 3 (2002)
  [arXiv:hep-th/0206255].

\bibitem{Dijkgraaf:2002dh}
  R.~Dijkgraaf and C.~Vafa,
  arXiv:hep-th/0208048.

\bibitem{Dijkgraaf:2009pc}
  R.~Dijkgraaf and C.~Vafa,
  arXiv:0909.2453 [hep-th].
\bibitem{Morales:1996bp}
  J.~F.~Morales and M.~Serone,
  Nucl.\ Phys.\  B {\bf 481}, 389 (1996)
  [arXiv:hep-th/9607193].

\bibitem{deRoo:1980mm}
  M.~de Roo, J.~W.~van Holten, B.~de Wit and A.~Van Proeyen,
  Nucl.\ Phys.\  B {\bf 173}, 175 (1980).
\bibitem{Nakayama:2010ff}
  Y.~Nakayama,
  JHEP {\bf 1007}, 054 (2010)
  [arXiv:1004.2986 [hep-th]].
\bibitem{Vafa:2000wi}
  C.~Vafa,
  J.\ Math.\ Phys.\  {\bf 42}, 2798 (2001)
  [arXiv:hep-th/0008142].
\bibitem{Beasley:2004ys}
  C.~Beasley and E.~Witten,
  JHEP {\bf 0501}, 056 (2005)
  [arXiv:hep-th/0409149].
\bibitem{Berkovits:2003pq}
  N.~Berkovits, H.~Ooguri and C.~Vafa,
  Commun.\ Math.\ Phys.\  {\bf 252}, 259 (2004)
  [arXiv:hep-th/0310118].
\bibitem{Gross:1990ub}
  D.~J.~Gross and I.~R.~Klebanov,
  Nucl.\ Phys.\  B {\bf 344}, 475 (1990).
\bibitem{Nakayama:2004vk}
  Y.~Nakayama,
  Int.\ J.\ Mod.\ Phys.\  A {\bf 19}, 2771 (2004)
  [arXiv:hep-th/0402009].
\bibitem{Sulkowski:2009ne}
  P.~Sulkowski,
  JHEP {\bf 1004}, 063 (2010)
  [arXiv:0912.5476 [hep-th]].

\bibitem{Nakayama:2007cf}
  Y.~Nakayama, M.~Taki, T.~Watari and T.~T.~Yanagida,
  Phys.\ Lett.\  B {\bf 655}, 58 (2007)
  [arXiv:0705.0865 [hep-ph]].

\bibitem{Nakayama:2007je}
  Y.~Nakayama,
  JHEP {\bf 0802}, 013 (2008)
  [arXiv:0712.0619 [hep-ph]].
\bibitem{Meade:2008wd}
  P.~Meade, N.~Seiberg and D.~Shih,
  Prog.\ Theor.\ Phys.\ Suppl.\  {\bf 177}, 143 (2009)
  [arXiv:0801.3278 [hep-ph]].

\end{thebibliography}
\end{document}